\documentclass[epj]{svjour}
\usepackage{amscd}
\usepackage{amssymb}
\usepackage{amsmath}
\usepackage{dcolumn}
\usepackage[T1]{fontenc}
\usepackage[latin1]{inputenc}
\usepackage{graphicx}
\usepackage{graphics}
\usepackage{color}
\def\be{\begin{equation}}
\def\ee{\end{equation}}
\def\ba{\begin{eqnarray}}
\def\ea{\end{eqnarray}}
\def\>{\rangle}
\def\<{\langle}
\def\n{\nonumber}

\begin{document}
\title{An analytic study of the ionization from an ultrathin quantum well
       in a weak electrostatic field}
\author{ILki Kim \thanks{\emph{e-mail}: hannibal.ikim@gmail.com}}
\institute{Department of Physics, North Carolina Central University,
           Durham, NC 27707, U.S.A.}
\date{\today}
\abstract{We consider the time evolution of a particle bound by an
attractive one-dimensional delta-function potential (at $x = 0$)
when a uniform electrostatic field ($F$) is applied. We explore
explicit expressions for the time-dependent wavefunction
$\psi_F(x,t)$ and the ionization probability ${\mathcal{P}}(t)$,
respectively, in the weak-field limit. In doing so, $\psi_F(0,t)$ is
a key element to their evaluation. We obtain a closed expression for
$\psi_F(0,t)$ which is an excellent approximation of the exact
result being a numerical solution of the Lippmann-Schwinger integral
equation. The resulting probability density $|\psi_F(0,t)|^2$, as a
simple alternative to ${\mathcal{P}}(t)$, is also in good agreement
to its counterpart from the exact one. In doing this, we also find a
new and useful integral identity of the Airy function.
}
\maketitle
\section{Introduction}
\label{intro} The ionization of atoms in an external electric field
is one of the oldest issues in quantum mechanics. As a simple
example, hydrogen-like atoms in a uniform electrostatic field have
extensively been considered \cite{YAM77}. Here, the background
potential caused by the field decreases without limit in one
direction, and electrons initially in the bound state will
eventually tunnel through the ``barrier'' created by the field, and
will ionize. Accordingly, there are no true bound states. The
tunneling rate for an ensemble of many independent electrons has
been calculated based on the exponential decay law following from
the experimentally supported statistical assumption that the
tunneling rate is proportional to the number of available atoms. On
the other hand, big experimental advances in the field of
nano-scaled physics have increased importance of the study of the
tunneling process of {\em individual} electrons; new nano-scaled
devices have been devised, examples of which are tunnel junctions
based on the electron resonant-tunneling effect \cite{JON89} and
molecular switches \cite{JOA86}. Therefore, a detailed understanding
of the capabilities of these devices clearly requires a deeper
knowledge of the tunneling process of a single electron subjected to
an external field. However, in analytic studies of the time
evolution of the tunneling process, e.g., leading to ionization, we
have the mathematical difficulty that there are no explicitly
solvable models for a transition from a bound state to the
continuum. Also, even obtaining the numerical solution with high
accuracy to this problem would not be an easy task either,
especially in the strong-field limit where a highly oscillatory
behavior is found in the time evolution of the bound-continuum
transition.

In this paper, we would like to study the time evolution of a
particle bound by an attractive one-dimensional delta-function
potential (at $x = 0$) when an external field is applied. We will
restrict our discussion below to a uniform electrostatic field $F$
for simplicity's sake. The single delta-function potential well
(without an external field) has a single bound state. This would
make easier an analysis of the ionization. From an applied point of
view, the delta-function potential system has actually been utilized
heuristically to represent a short range atom or optically active
defect as well as, in more complex combinations, resonant-tunneling
junctions \cite{BEN85,SHN97,DAT97}, molecular switches
\cite{DAT97,COL99,LIU03}, and Dirac comb lattices \cite{KIA74}. This
model of the field-induced time-dependent ionization was first
discussed by Geltman \cite{GEL77}. Later on, some various approaches
to explicitly obtaining the time-dependent solution $\psi_F(x,t)$
have been carried out \cite{LUD87,ELK88,SUS90,KLE94,ENG95,ROK00}.
However, no exact solution in closed form (in terms of its actual
calculability) has been found. The method we adopt here to solve the
time-dependent Schr\"{o}dinger equation is to turn it into an
integral equation based on the Lippmann-Schwinger formalism. The
integral equation has thus far been focused mainly upon its
numerical solvability. Also, we will employ both the $\hat{x}\,F$
interaction Hamiltonian (the scalar-potential gauge) and the
$\hat{p}\,A$ (the vector-potential gauge) and then compare them
({\em cf.} for a detailed discussion of $\hat{x}\,F$ versus
$\hat{p}\,A$ gauge problem, see Ref. \cite{SCH84}).

A primary goal of this work is to pursue the analytical expressions
for the wavefunction $\psi_F(x,t)$ and the ionization probability
${\mathcal{P}}(t)$, respectively, in the weak-field limit ({\em
i.e.,} the field strength $f \lesssim 1$ relative to the strength of
the potential well). As will be shown, $\psi_F(0,t)$ is a key
ingredient to doing so. In \cite{ELK88} and \cite{ELB87}, on the
other hand, the ionization probabilities in the strong-field limit
have been studied while neglecting the influence of the residual
zero-range potential on the wavefunction $\psi_F(x,t)$ and using the
numerical method in the Born-approximation scheme, respectively.
They have been then compared to the exact result obtained from the
numerical analysis. In addition, the ionization probability in the
strong-field limit obtained in the scheme of the exponential decay
approximation was shown to be a good approximation on the average to
the exact result, although it cannot account for the short-time
ripples found in the exact result \cite{ELK88}. In the weak-field
limit, however, we cannot easily neglect the influence of the
ultrathin potential well to the wavefunction. We will obtain a
closed expression for $\psi_F(0,t)$ in this limit, and study its
probability density $|\psi_F(0,t)|^2$ as a simple alternative to
${\mathcal{P}}(t)$.

The general layout of this paper is the following; in Sect.
\ref{sec:review} we briefly review the known results including the
delta-function potential problem without an external field and the
problem of a particle subjected to a uniform electrostatic field but
not bound by the ultrathin potential well. Here, we also derive the
Lippmann-Schwinger integral equation for the current problem from
the time-dependent Schr\"{o}dinger equation, and review the
time-dependent tunneling through the ultrathin potential well ({\em
or} barrier) in the field-free case, which is an analytically
solvable model \cite{ELB88}. In Sect.
\ref{sec:semiclassical_wavefunction}, analytical expressions for
$\psi_F(x,t)$ and ${\mathcal{P}}(t)$ are explored, and an explicit
expression for $\psi_F(0,t)$ in the weak-field limit is derived,
partially with the aid of the exponential decay approximation.
Section \ref{sec:ionization_probability} deals with
$|\psi_F(0,t)|^2$, as a simple alternative to the ionization
probability, which is in excellent agreement to the exact result
obtained from the numerical analysis of the Lippmann-Schwinger
equation; this contains the slack ripples in the time evolution,
which cannot be found in its counterpart obtained entirely from the
exponential decay approximation. Finally, in Sect.
\ref{sec:conclusion} we give the conclusion of this paper.

\section{General formulation}\label{sec:review}
The system under consideration is described by the Hamiltonian
\begin{equation}\label{eq:total_hamiltonian1}
    \hat{H}\; =\; {\textstyle \frac{\hat{p}^2}{2 m}\, -\, V_0\,\delta(\hat{x})\, -\, \hat{x}\,F(t)\,,}
\end{equation}
where $V_0 > 0$ and $F(t) = F \cdot \Theta(t)$. We decompose this
Hamiltonian into $\hat{H}_0 = \frac{\hat{p}^2}{2 \mu} -
V_0\,\delta(\hat{x})$ and $\hat{H}_F = \frac{\hat{p}^2}{2 \mu} -
\hat{x}\,F$, where $\mu = 2 m$ and $\hat{H} = \hat{H}_0\, +\,
\hat{H}_F$. First, $\hat{H}_0$ has a single bound state,
\begin{equation}\label{eq:bound_state_H0}
    {\textstyle \psi_b(x)\; =\; \sqrt{B}\; e^{-B\,|x|}\,,}
\end{equation}
with eigen energy $E_b = -\frac{\hbar^2 B^2}{2 \mu}$, where $B =
\frac{\mu V_0}{\hbar^2}$. Clearly, $\psi_b^{\prime}(x)$ has a
discontinuity at $x = 0$. All eigenstates and eigenvalues of
$\hat{H}_0$, and also the completeness of the eigenstates, were
discussed in detail in, e.g., Refs. \cite{DAM75} and \cite{GOT03}.
The eigenfunction of $\hat{H}_F$ with (continuous) energy $E$ is
given by \cite{VAL04}
\begin{equation}\label{eq:eigenfunction_of_hamiltonian_f}
    {\textstyle \phi_E(x)\; =\; \left(\frac{4\,\mu^2}{\hbar^4\,F}\right)^{\frac{1}{6}}\;
    \text{Ai}\left\{-\left(\frac{2\,\mu\,F}{\hbar^2}\right)^{\frac{1}{3}}
    \left(x + \frac{E}{F}\right)\right\}\,,}
\end{equation}
where the Airy function
\begin{equation}\label{eq:airy_function_def1}
    {\textstyle \text{Ai}(\sigma)\; :=\; \frac{1}{2\pi}\,
    \int_{-\infty}^{\infty}\,d y\; e^{i \frac{y^3}{3}\, +\, i \sigma
    y}\,.}
\end{equation}
We have an alternative choice to $\hat{x}\,F$ in equation
(\ref{eq:total_hamiltonian1}) for the field-interaction Hamiltonian;
the Hamiltonian $\hat{H}$ in the scalar-potential gauge is then
replaced by its counterpart in the vector-potential gauge,
\begin{equation}\label{eq:total_hamiltonian2}
    \hat{H}_v\; =\; {\textstyle \hat{H}_{v,F}\; -\; V_0\,\delta(\hat{x})\,,}
\end{equation}
where $\hat{H}_{v,F}\, =\, \frac{1}{2 m} \left\{\hat{p}\, +\,
p_c(t)\right\}^2$ with the vector potential $p_c(t) = \int_0^t
F(\tau)\,d\tau$. The time-dependent Schr\"{o}dinger equations in the
scalar-potential and the vector-potential gauges read
\begin{eqnarray}\label{eq:time_dependent_Schroedinger}
    {\textstyle i \hbar\,\frac{\partial}{\partial t}\, \psi_F(x,t)}
    &=& {\textstyle \hat{H}\, \psi_F(x,t)\,;}\n\\
    {\textstyle i \hbar\,\frac{\partial}{\partial t}\, \psi_v(x,t)}
    &=& {\textstyle \hat{H}_v\, \psi_v(x,t)\,,}
\end{eqnarray}
respectively. Here, the two equations are connected by the
relationship $\psi_F(x,t) = e^{\frac{i}{\hbar} x \cdot p_c(t)}\,
\psi_v(x,t)$. Remarkably enough, the ionization probability
${\mathcal{P}}(\tau) = 1 - |\<\psi_b|\psi_F(\tau)\>|^2$ in the
scalar-potential gauge is different from ${\mathcal{P}}_v(\tau) = 1
- |\<\psi_b\cdot e^{-\frac{i}{\hbar} p_c(\tau)\cdot
x}|\psi_F(\tau)\>|^2$ in the vector potential gauge. As was pointed
out in \cite{ELB87}, for ${\mathcal{P}}(\tau)$ one switches on the
field at $t=0$ and turns it off at $t=\tau$. Afterward the
ionization probability is measured; for ${\mathcal{P}}_v(\tau)$ one
turns off the vector potential instead of the field. However, it
appears physically unrealistic to think of an experiment where the
vector potential $p_c(\tau) = \int_0^{\tau}
F(\tau^{\prime})\,d\tau^{\prime}$ is turned off. Therefore we will
take ${\mathcal{P}}(\tau)$ under consideration.

We intend to derive the Lippmann-Schwinger integral equation for
this problem from the time-dependent Schr\"{o}dinger equation. To do
this, let us go ahead with the Schr\"{o}dinger equation in the
vector-potential gauge in the momentum representation, where the
homogeneous solution (to the Schr\"{o}dinger equation for
$\hat{H}_{v,F}$) can be obtained very easily. By substituting
\begin{equation}\label{eq:x_representation}
    \psi_v(x,t)\; =\; {\textstyle \frac{1}{\sqrt{2 \pi \hbar}}\,
    \int_{-\infty}^{\infty}\, dp\; \varphi_v(p,t)\; e^{\frac{i}{\hbar}
    x p}}
\end{equation}
into the equation for $\psi_v(x,t)$ in
(\ref{eq:time_dependent_Schroedinger}) and then multiplying
$\frac{1}{2 \pi \hbar}\,e^{-\frac{i}{\hbar}x p'}$ on both sides,
followed by the integration over $x$, we acquire
\begin{eqnarray}\label{eq:schroedinger_eq_in_p}
    {\textstyle i \hbar\,\frac{\partial}{\partial t}\,\varphi_v(p,t)}
    &=& {\textstyle \frac{\left\{p\, +\, p_c(t)\right\}^2}{2\,m}\,
    \varphi_v(p,t)\; +}\n\\
    && {\textstyle \int_{-\infty}^{\infty} dp'\, \tilde{V}(p-p')\, \varphi_v(p',t)\,,}
\end{eqnarray}
where
\begin{eqnarray}
    {\textstyle \tilde{V}(p)} &=& {\textstyle \frac{1}{2 \pi
    \hbar}\,\int_{-\infty}^{\infty}\, dx\; e^{-\frac{i}{\hbar} p\,x}\,
    V(x)\,,}\n\\
    {\textstyle V(x)} &=& {\textstyle
    \int_{-\infty}^{\infty}\, dp\; e^{\frac{i}{\hbar} p\,x}\,
    \tilde{V}(p)\, =\, -V_0\,\delta(x)\,.}
\end{eqnarray}

First, the (homogeneous) solution to the equation,\\${\textstyle i
\hbar\,\frac{\partial}{\partial t}\,\varphi_{v,F}(p,t)}\, =\,
{\textstyle \frac{\left\{p\, +\, p_c(t)\right\}^2}{2\,m}\,
\varphi_{v,F}(p,t)}$ easily appears as
\begin{equation}\label{eq:homogeneous_sol}
    {\textstyle \varphi_{v,F}(p,t)\; =\; U_{v,F}(p,t) \cdot \varphi(p,0)\,,}
\end{equation}
where the time evolution factor\\$U_{v,F}(p,t)\, =\,
e^{-\frac{i}{\hbar} \int_0^{t} \frac{1}{2 m} \{p\, +\,
p_c(\tau)\}^2\, d\tau}$, and the initial bound state
\begin{equation}\label{eq:bound_state_momentum_rep}
    {\textstyle \varphi(p,0)\; =\; \varphi_b(p)\; =\; \textstyle
    \sqrt{\frac{2}{\pi}}\, \frac{(\hbar\,B)^{3/2}}{p^2\, +\,
    (\hbar\,B)^2}\,.}
\end{equation}
Next, by inserting an ansatz $\varphi_v(p,t) = \varphi_{v,F}(p,t)\,
\chi_v(p,t)$ with $\chi_v(p,0) = 1$ into equation
(\ref{eq:schroedinger_eq_in_p}), we have
\begin{equation}\label{eq:schroedinger_in_p_for_chi}
    {\textstyle i \hbar\,\frac{\partial}{\partial t}\,\chi_v(p,t)}\;
    =\; {\textstyle \frac{1}{2 \pi \hbar}\,\frac{-V_0}{\varphi_{v,F}(p,t)}\,
    \int_{-\infty}^{\infty} dp'\, \varphi_{v,F}(p',t)\,
    \chi_v(p',t)\,.}
\end{equation}
Integrating equation (\ref{eq:schroedinger_in_p_for_chi}) over $t$
and then multiplying\\$\varphi_{v,F}(p,t)$, we obtain
\begin{eqnarray}\label{eq:sol_schroedinger_eq_in_p}
    \varphi_v(p,t) &=& {\textstyle \varphi_{v,F}(p,t)\, +\,
    \frac{i}{\hbar}\, \frac{V_0}{2 \pi \hbar}\,
    U_{v,F}(p,t)\; \times}\n\\
    && {\textstyle \int_0^t d\tau\; U_{v,F}^{-1}(p,\tau)\, \int_{-\infty}^{\infty} dp'\,
    \varphi_v(p',\tau)\,.}
\end{eqnarray}
By using the Fourier transform in equation
(\ref{eq:x_representation}), we arrive at the the Lippmann-Schwinger
equation for $\psi_v(x,t)$, which reads
\begin{equation}\label{eq:sol_schroedinger_eq_in_x}
    {\textstyle \psi_v(x,t)\, =\, \phi_{v}(x,t)\, +\,
    \frac{i}{\hbar}\,V_0\, \int_0^t\, d\tau\, K_{v,F}(x,t|0,\tau)\, \psi_v(0,\tau)\,.}
\end{equation}
Here, the homogeneous solution\\$\phi_{v}(x,t)\, =\, {\textstyle
\frac{1}{\sqrt{2 \pi \hbar}}\, \int_{-\infty}^{\infty}\, dp\;
\varphi_{v,F}(x,t)\, e^{\frac{i}{\hbar} x p}}$, and the propagator
$K_{v,F}(x,t|x',\tau)$ appears as the Fourier transform of
$U_{v,F}(p,t)\, U_{v,F}^{-1}(p,\tau)$ such that
\begin{eqnarray}\label{eq:kernel1}
    \hspace*{-.7cm}&& {\textstyle K_{v,F}(x,t|x',\tau)\; =}\\
    \hspace*{-.7cm}&& {\textstyle \frac{1}{2 \pi \hbar}\, \int_{-\infty}^{\infty}\,
    dp \cdot e^{\frac{i}{\hbar} p\,(x - x')} \cdot U_{v,F}(p,t) \cdot
    U_{v,F}^{-1}(p,\tau)\; =}\n\\
    \hspace*{-.7cm}&& {\textstyle \sqrt{\frac{m}{2 \pi i\,\hbar\,(t - \tau)}} \cdot e^{-\frac{i}{\hbar}
    \{S_c(t) - S_c(\tau)\}} \cdot e^{\frac{i}{\hbar} \frac{m}{2\,(t - \tau)} (\{x -
    x_c(t)\} - \{x' - x_c(\tau)\})^2}}\n
\end{eqnarray}
with the field-induced translation\\$x_c(t)\, =\, {\textstyle
\frac{1}{m}\, \int_0^t\, d\tau\, p_c(\tau)\, =\,
\frac{F\,t^2}{2\,m}}$\, and the field-induced action $S_c(t)\, =\,
{\textstyle \frac{1}{2\,m} \int_0^t\, d\tau\, p_c^2(\tau)}$
${\textstyle =\, \frac{F^2\,t^3}{6\,m}}$. Then, it turns out that
$\phi_{v}(x,t)\, =\, \int_{-\infty}^{\infty} d
x'\,K_{v,F}(x,t|x',0)\,\psi(x',0)$, where\\$\psi(x',0) =
\psi_b(x')$. On the right hand side of equation
(\ref{eq:sol_schroedinger_eq_in_x}), this clearly represents free
motion subjected to an external field (so-called Volkov part), and
the second term is the influence of the residual zero-range
potential.

Keeping in mind the gauge factor $e^{\frac{i}{\hbar} x \cdot
p_c(t)}$, we then easily find the propagator $K_F(x,t|x',\tau)$ in
the\\scalar-potential gauge, from (\ref{eq:kernel1}) for the
vector-potential gauge, as
\begin{equation}\label{eq:kernel1_s}
    {\textstyle K_F(x,t|x',\tau)}\; =\; {\textstyle K_{v,F}(x,t|x',\tau) \cdot e^{\frac{i}{\hbar} \{x p_c(t)\, -\, x'
    p_c(\tau)\}}\,,}
\end{equation}
and also the corresponding Lippmann-Schwinger equation
\begin{equation}\label{eq:sol_schroedinger_eq_in_x_dc1}
    {\textstyle \psi_F(x,t)\; =\; \phi_F(x,t)\, +\,
    \frac{i}{\hbar}\,V_0\,\int_0^t\, d\tau\, K_F(x,t|0,\tau) \cdot \psi_F(0,\tau)\,.}
\end{equation}
Equivalently, we have the integral equation for the total propagator
\begin{eqnarray}\label{eq:sol_schroedinger_eq_in_x_dc1_k}
    {\textstyle \mathcal{K}_F(x,t|x',0)} &=& {\textstyle K_F(x,t|x',0)\, +\,
    \frac{i}{\hbar}\,V_0\; \times}\\
    && {\textstyle \int_0^t\, d\tau\, K_F(x,t|0,\tau) \cdot
    \mathcal{K}_F(0,\tau|x',0)\,.}\n
\end{eqnarray}
Here, the homogeneous solution\\$\phi_F(x,t)\, =\,
\int_{-\infty}^{\infty} d x'\,K_F(x,t|x',0)\,\psi_b(x')$ with
\begin{equation}\label{eq:kernel2_s}
    {\textstyle K_F(x,t|x',0)}\; =\; {\textstyle \sqrt{\frac{m}{2 \pi i\,\hbar t}} \cdot
    e^{\frac{i m (x - x')^2}{2 \hbar t}} \cdot
    e^{-\frac{i}{\hbar}\left(\frac{F^2}{24 m}\,t^3 - \frac{x + x'}{2} F\,t\right)}}
\end{equation}
reduces to a closed expression \cite{ELK88}
\begin{eqnarray}\label{eq:homog_sol1}
    \phi_F(x,t) &=& {\textstyle \sqrt{B}\; e^{\frac{i}{\hbar}
    \left\{x p_c(t)\, -\, S_c(t)\right\}}\, \left\{M\left(x - x_c(t); -iB;
    \frac{\hbar}{m}t\right)\right.}\n\\
    && {\textstyle \left. +\, M\left(x_c(t) - x; -iB;
    \frac{\hbar}{m}t\right)\right\}}
\end{eqnarray}
in terms of the Moshinsky function \cite{MOS52}
\begin{equation}\label{eq:moshinsky_function}
    {\textstyle M(x;k;t)\; =\; \frac{1}{2}\; e^{i\,\left(k\,x\, -\, \frac{1}{2}\,k^2\,t\right)} \cdot
    \mbox{erfc}\left\{\frac{x\, -\, k\,t}{\sqrt{2\,i\,t}}\right\}\,,}
\end{equation}
where $\text{erfc}(z)$ is the complementary error function. Also, we
have in equation (\ref{eq:sol_schroedinger_eq_in_x_dc1})
\begin{equation}\label{eq:kernel1_dc}
    {\textstyle K_F(x,t|0,\tau)}\; =\; {\textstyle K_0(x,t|0,\tau) \cdot e^{\frac{i}{\hbar}
    \{\frac{F x}{2} (t - \tau)\, -\, \frac{F^2}{24 m} (t - \tau)^3\}}\,,}
\end{equation}
where the field-free propagator
\begin{equation}\label{eq:field_free_propagator}
    {\textstyle K_0(x,t|0,\tau)}\; =\; {\textstyle \sqrt{\frac{m}{2 \pi i\,\hbar\,(t - \tau)}} \cdot
    e^{\frac{i}{\hbar} \frac{m}{2\,(t - \tau)}\,x^2}\,.}
\end{equation}

It is also interesting to consider the momentum representation of
the propagator $\mathcal{K}_F(x,t|x',0)$ in equation
(\ref{eq:sol_schroedinger_eq_in_x_dc1_k}),
\begin{eqnarray}\label{eq:momentum_representaion_k_1}
    {\textstyle \<p|\hat{U}(t)|p'\>} &=& {\textstyle
    \tilde{\mathcal{K}}_F(p,t|p',0)\; =}\\
    && {\textstyle \int_{-\infty}^{\infty} dx\, \int_{-\infty}^{\infty} dx'\,
    \<p|x\>\, \mathcal{K}_F(x,t|x',0)\, \<x'|p'\>\,,}\n
\end{eqnarray}
where $\hat{U}(t) = e^{-\frac{i}{\hbar} \hat{H} t}$ and $\<p|x\> =
e^{-\frac{i}{\hbar} x p}/\sqrt{2 \pi \hbar}$\,. By applying an
iteration of replacing the integrand $\mathcal{K}_F(0,\tau|x',0)$ by
the entire expression of the right hand side and then performing the
Fourier transform to each term, we will obtain an explicit
expression for $\tilde{\mathcal{K}}_F(p,t|p',0)$; after a fairly
lengthy calculation we eventually arrive at the expression
\begin{equation}\label{eq:momentum_representaion_k_2}
    {\textstyle \tilde{\mathcal{K}}_F(p,t|p',0)\; =\; U_{v,F}(p-p_c(t), t)\;
    \mathcal{A}_F(p,p',t)\,,}
\end{equation}
where
\begin{eqnarray}\label{eq:momentum_representaion_k_3}
    \hspace*{-.5cm}&&\mathcal{A}_F(p,p',t)\; =\; {\textstyle \delta(p-p_c(t)-p')\,
    +\, \int_0^t d\tau_1\, U_{v,F}^{\ast}(p-p_c(t),\tau_1)\, \times}\n\\
    \hspace*{-.5cm}&& \left\{\,\sum_{k=1}^{\infty} {\textstyle (\frac{i}{2 \pi \hbar^2}
    V_0)^{k}}\,
    \prod_{l=1}^{k-1} {\textstyle \int_0^{\tau_l} d\tau_{l+1}
    \int_{-\infty}^{\infty} dp_l\; U_{v,F}(p_l,\tau_{l})\;
    U_{v,F}^{\ast}(p_l,\tau_{l+1})}\,\right\}\n\\
    \hspace*{-.5cm}&& {\textstyle \times\, U_{v,F}(p',\tau_{k})\,.}
\end{eqnarray}
Here, the summation index $k$ counts how many times the electron
interacts with the delta-well while it travels from $p$ to $p'$. The
Born term ($k = 1$) has a closed form
\begin{eqnarray}\label{eq:born_term_momentum_rep}
    {\textstyle \mathcal{A}_{F}^{(B)}(p,p',t)} &=&
    {\textstyle \frac{i}{2 \hbar} V_0
    \sqrt{\frac{i m}{2 \pi \hbar\,F\,\{p - p' - p_c(t)\}}}\; \times}\\
    && {\textstyle e^{-\frac{i}{8 \hbar m F}
    \{p - p' - p_c(t)\} \{p + p' - p_c(t)\}^2}\; \times}\n\\
    && {\textstyle \left\{\mbox{erfc}\left(\frac{b_F}{\sqrt{a_F}}\right)\, -\,
    \mbox{erfc}\left(\sqrt{a_F}\,
    t\,+\,\frac{b_F}{\sqrt{a_F}}\right)\right\}\,.}\n
\end{eqnarray}
where $a_F(t) = \frac{F}{2 i \hbar m} \{p - p' - p_c(t)\}$ and\,
$b_F(t) = \frac{1}{4 i \hbar m} \{(p - p_c(t))^2 - (p')^2\}$. This
momentum representation $\tilde{\mathcal{K}}_F(p,t|p',0)$ in the
scalar-potential gauge in terms of $\{U_{v,F}(p,t)\}$ from the
vector-potential gauge rather straightforwardly demonstrates the
complexity of the bound-continuum transition involving infinitely
many complicated continuum-continuum transitions.

Now, we briefly consider the field-free ($F=0$) case described by
the Hamiltonian $\hat{H}_0$. In this case, we can obtain an explicit
expression for $\psi_{F=0}(x,t)$ from the Lippmann-Schwinger
equation (\ref{eq:sol_schroedinger_eq_in_x_dc1}) \cite{ELB88}. We
sketch this procedure; by applying the Laplace transform to this
equation for $x=0$, with the aid of the convolution theorem, we
obtain
\begin{equation}\label{eq:laplace_psi_x0_a.c}
    {\textstyle \tilde{\psi}_0(0,s)\; =\; \frac{\tilde{\phi}_0(0,s)}{1\, -\,
    \frac{i}{\hbar} \sqrt{\frac{m}{2 i \hbar\,s}}\; V_0}\,,}
\end{equation}
where the Laplace transformed $\tilde{\psi}_0(x,s)\, =\,
\mathcal{L}\left\{\psi_0(x,t)\right\}$, and $\tilde{\phi}_0(x,s)\,
=\, \mathcal{L}\left\{\phi_0(x,t)\right\}$. From equation
(\ref{eq:homog_sol1}) for $F = 0$ and $x = 0$, it follows that
\begin{equation}\label{eq:laplace_psi_x0_f0}
    {\textstyle \phi_0(0,t)\; =\; \sqrt{B}\; e^{-\frac{i}{\hbar} E_b t}\;
    \mbox{erfc}\left(\sqrt{-\frac{i}{\hbar}\,E_b\,t}\right)\,,}
\end{equation}
which gives ${\textstyle \tilde{\phi}_0(0,s)}\, =\, {\textstyle
\frac{\sqrt{B}}{\sqrt{s}\, \left(\sqrt{s}\, +\,
\sqrt{-\frac{i}{\hbar}\,E_b}\right)}}$. Applying the inverse Laplace
transform to equation (\ref{eq:laplace_psi_x0_a.c}) \cite{ROB66}, we
now acquire $\psi_0(0,t)\, =\, \sqrt{B}\, e^{-\frac{i}{\hbar}
E_b\,t}$, which subsequently yields, from equation
(\ref{eq:sol_schroedinger_eq_in_x_dc1}), \cite{ABR74}
\begin{eqnarray}\label{eq:wave_function_without_field}
    &&{\textstyle \psi_0(x,t)}\; =\n\\
    &&{\textstyle \phi_0(x,t)\, +\,
    \sqrt{B}\,\left\{M\left(|x|; iB;\frac{\hbar}{m}t\right)\, -\,
    M\left(|x|; -iB; \frac{\hbar}{m}t\right)\right\}\; =}\n\\
    && {\textstyle \sqrt{B}\,\left\{M\left(|x|; iB;\frac{\hbar}{m}t\right)\, +\,
    M\left(-|x|; -iB; \frac{\hbar}{m}t\right)\right\}\; =}\n\\
    && {\textstyle \psi_b(x)\; e^{-\frac{i}{\hbar} E_b t}\,.}
\end{eqnarray}
Along the same line, even the total propagator
$\mathcal{K}_0(x,t|x',0)$ for $\hat{H}_0$ was exactly derived in
Ref.~\cite{ELB88}. On the other hand it is very non-trivial to apply
the Laplace transform to equation
(\ref{eq:sol_schroedinger_eq_in_x_dc1}) for $F \ne 0$ [note the
highly oscillatory factor $e^{-\frac{F^2}{24 m} (t - \tau)^3}$ in
equation (\ref{eq:kernel1_dc})] in order to derive the total
propagator for $\hat{H}$, the explicit expression of which has thus
far not been known. In fact, equation
(\ref{eq:sol_schroedinger_eq_in_x_dc1}) has been just numerically
treated, namely, first it is numerically solved for $\psi_F(0,t)$,
which is substituted back into it for $\psi_F(x,t)$.

\section{Explicit expressions for wavefunction and
ionization probability}\label{sec:semiclassical_wavefunction}
%
In dealing with the the temporal integral in
(\ref{eq:sol_schroedinger_eq_in_x_dc1}) with (\ref{eq:kernel1_dc})
analytically, the factor $e^{-\frac{F^2}{24 m} (t - \tau)^3}$ in
$K_F(x,t|0,\tau)$ is a major limiting one. We intend to circumvent
this difficulty with the aid of the Fourier transform of Airy
function $\text{Ai}(\sigma)$ \cite{VAL04},
\begin{equation}\label{eq:airy_function_rel1}
    {\textstyle \int_{-\infty}^{\infty}\, d\sigma\, \text{Ai}(\sigma)\; e^{i \sigma
    \eta}\; =\; e^{-\frac{i}{3} \eta^3}\; =\; e^{-\frac{i}{\hbar} \frac{F^2}{24 m} (t -
    \tau)^3}}
\end{equation}
where the dimensionless ``time'' $\eta = \frac{1}{2}
(\frac{F^2}{\hbar\,m})^{\frac{1}{3}}\,(t - \tau)$ \cite{KIM06}, and
the dimensionless ``energy'' $\sigma$. Therefore, we have, from
equation (\ref{eq:kernel1_dc}),
\begin{eqnarray}\label{eq:kernel1_dc1}
    {\textstyle K_F(x,t|0,\tau)} &=&
    {\textstyle K_0(x,t|0,\tau)\; e^{\frac{i}{\hbar}
    \frac{F x}{2} (t - \tau)}\; \times}\n\\
    && {\textstyle \int_{-\infty}^{\infty} d\sigma\,
    \text{Ai}(\sigma)\; e^{\frac{i}{2} \sigma
    (\frac{F^2}{\hbar\,m})^{\frac{1}{3}} (t - \tau)}\,.}
\end{eqnarray}
By means of the eigenfunctions $\phi_E(x=0)$ in equation
(\ref{eq:eigenfunction_of_hamiltonian_f}) and the substitution
$\sigma = - \{\frac{2 m}{(\hbar\,F)^2}\}^{\frac{1}{3}}\, E$, this
becomes
\begin{eqnarray}\label{eq:kernel1_dc2}
    {\textstyle K_F(x,t|0,\tau)} &=& {\textstyle \int_{-\infty}^{\infty} d E\;
    K_0(x,t|0,\tau)\; \frac{1}{\sqrt{F}}\; \times}\n\\
    && {\textstyle e^{\frac{i}{\hbar}
    \left(\frac{F x}{2}\,-\,\frac{E}{2^{2/3}}\right)\,(t - \tau)}\; \phi_E(0)\,.}
\end{eqnarray}
Here, we see that each $\sigma$ may be interpreted as a spectrum
channel of the propagator $K_F(x,t|0,\tau)$. From equations
(\ref{eq:sol_schroedinger_eq_in_x_dc1}) and (\ref{eq:kernel1_dc1})
it follows that
\begin{equation}\label{eq:sol_schroedinger_eq_in_x_dc2}
    {\textstyle \psi_F(x,t)\; =\; \phi_F(x,t)\, +\,
    B\,\sqrt{\frac{i \hbar}{2 \pi m}}\,
    \int_{-\infty}^{\infty} d\sigma\, \text{Ai}(\sigma)\; T_F(x,t,\sigma)\,,}
\end{equation}
where
\begin{equation}\label{eq:particular_sol_schroedinger_eq_in_x_dc2_2}
    {\textstyle T_F(x,t,\sigma)\; :=\; \int_0^t d \tau\, \frac{1}{\sqrt{t - \tau}}\; e^{\frac{i}{\hbar} \frac{m\,x^2}{2\,(t - \tau)}}
    \cdot e^{\frac{i}{\hbar}\,\varepsilon_F(x,\sigma)\, (t - \tau)} \cdot \psi_F(0,\tau)}
\end{equation}
with the field-induced energy $\varepsilon_F(x,\sigma) = \frac{F
x}{2}\, +\, \frac{\sigma}{2} (\frac{\hbar^2 F^2}{m})^{\frac{1}{3}}$.
We also note that the eigen energy $E =
-2^{\frac{2}{3}}\,\varepsilon_F(0,\sigma)$. To explicitly evaluate
the integration over $\sigma$ in
(\ref{eq:sol_schroedinger_eq_in_x_dc2}), we expand
$\text{Ai}(\sigma)$ in terms of the delta-function $\delta(\sigma)$
so that
\begin{equation}\label{eq:airy_function_expansion1}
    {\textstyle \text{Ai}(\sigma)\; =\;} \sum_{k=0}^{\infty} {\textstyle \frac{(-1)^k}{k!\; 3^k}
    \left(\frac{\partial}{\partial \sigma}\right)^{3k}\,
    \delta(\sigma)\,,}
\end{equation}
which follows from the definition of the Airy function in equation
(\ref{eq:airy_function_def1}). By means of the relationship $\int
d\sigma\; \delta^{(k)}(\sigma)\; h(\sigma) = (-1)^k\, h^{(k)}(0)$,
equation (\ref{eq:sol_schroedinger_eq_in_x_dc2}) becomes
\begin{eqnarray}\label{eq:airy_representation}
    {\textstyle \psi_F(x,t)} &=& {\textstyle \phi_F(x,t)\, +\,
    B\,\sqrt{\frac{i \hbar}{2 \pi m}}\; \times}\n\\
    && \sum_{k=0}^{\infty}\, {\textstyle \frac{1}{k!\; 3^k}\; \left.\frac{\partial^{3k}}{\partial \sigma^{3k}}\,
    T_F(x,t,\sigma)\right|_{\sigma=0}\,.}
\end{eqnarray}

We now calculate $\<\psi_b|\psi_F(t)\>$ for the ionization
probability ${\mathcal{P}}(t) = 1 - |\<\psi_b|\psi_F(t)\>|^2$. From
equations (\ref{eq:bound_state_H0}),
(\ref{eq:particular_sol_schroedinger_eq_in_x_dc2_2}), and
(\ref{eq:airy_representation}), we obtain
\begin{eqnarray}\label{eq:ionization_prob_pre1}
    {\textstyle \left\<\psi_b|\psi_F(t)\right\>} &=&
    {\textstyle \left\<\psi_b|\phi_F(t)\right\>\, +\,
    \frac{i\,\hbar}{2\,m}\,\sqrt{B}^3}\; \times\n\\
    && \sum_{k=0}^{\infty}\, {\textstyle \frac{1}{k!\; 3^k}\, \left.\frac{\partial^{3k}}{\partial \sigma^{3k}}\,
    G_F(t,\sigma)\right|_{\sigma = 0}\;,}
\end{eqnarray}
where
\begin{eqnarray}\label{eq:ionization_prob_pre2}
    {\textstyle G_F(t,\sigma)} &=& {\textstyle \int_0^t d\tau\; \psi_F(0,\tau)\; e^{i \gamma_1(\sigma) \cdot (t -
    \tau)}\; \times}\n\\
    && {\textstyle \left\{e^{- \gamma_2\,(t - \tau)^2}\, \text{erfc}\hspace*{-.07cm}\left(-\gamma_3\,\sqrt{t - \tau}^3\,
    +\, \gamma_4\,\sqrt{t - \tau}\right)\; +\right.}\n\\
    && {\textstyle \left. e^{\gamma_2\,(t - \tau)^2}\, \text{erfc}\hspace*{-.07cm}\left(\gamma_3\,\sqrt{t - \tau}^3\, +\,
    \gamma_4\,\sqrt{t - \tau}\right)\right\}}
\end{eqnarray}
\hspace*{-.5cm}with $\gamma_1(\sigma)\, =\, \sigma
\left(\frac{F^2}{2\,\hbar\,m}\right)^{\frac{1}{3}} +
\frac{\hbar\,B^2}{2\,m}\,;\, \gamma_2\, =\, \frac{B F}{2\,m}\,;\,
\gamma_3\, =\, \frac{F}{2} \sqrt{\frac{1}{2\,i\,\hbar\,m}}$\,, and\,
$\gamma_4\, =\, B \sqrt{\frac{i\,\hbar}{2\,m}}\;$. Equation
(\ref{eq:ionization_prob_pre2}) yields
\begin{eqnarray}
    \hspace*{-1.5cm}&&{\textstyle \left.\frac{\partial^{3k}}{\partial \sigma^{3k}}\,
    G_F(t,\sigma)\right|_{\sigma = 0}}\; =\; {\textstyle \left(\frac{F^2}{2\,i\,\hbar\,m}\right)^k\, \int_0^t d\tau\, (t -
    \tau)^{3k}\; g_F(t,\tau)}\label{eq:ionization_prob_pre3}\\
    \hspace*{-1.5cm}&=& {\textstyle \left(\frac{F^2}{2\,i\,\hbar\,m}\right)^{k}
    (3k)!}\, \sum_{l=0}^{\infty}\,{\textstyle \frac{1}{(3k + l + 1)!}\, \left.\frac{\partial^l}{\partial
    \tau^l}\, g_F(t,\tau)\right|_{\tau = 0}\; t^{3k + l + 1}}\label{eq:ionization_prob_pre4}
\end{eqnarray}
where
\begin{eqnarray}\label{eq:ionization_prob_pre5}
    g_F(t,\tau) &=& {\textstyle 2\; \psi_F(0,\tau)\,
    \left\{M\hspace*{-.07cm}\left(-x_c(t - \tau); -i B;
    \frac{\hbar}{m}\,(t - \tau)\right)\right.}\n\\
    && {\textstyle \left.+\, M\hspace*{-.07cm}\left(x_c(t - \tau); -i B; \frac{\hbar}{m}\,(t -
    \tau)\right)\right\}}
\end{eqnarray}
in terms of the Moshinsky function $M(x; k; t)$. In obtaining
equation~(\ref{eq:ionization_prob_pre4}) from
(\ref{eq:ionization_prob_pre3}), an iteration of the integrations by
parts was applied. From equations (\ref{eq:ionization_prob_pre1}),
(\ref{eq:ionization_prob_pre4}) and (\ref{eq:ionization_prob_pre5})
we clearly see that the knowledge of $\psi_F(0,t)$ is the key to the
actual evaluation of the ionization probability. Therefore, we
restrict our discussion below to derive an explicit expression for
$\psi_F(0,t)$, a good approximation of which can actually be
obtained in the weak-field limit.

We will consider two different approximation schemes for
$\psi_F(0,t)$ in the weak-field limit; first, we simply replace
$K_F(x,t|0,\tau)$ and $\psi_F(0,\tau)$ on the right hand side of
equation (\ref{eq:sol_schroedinger_eq_in_x_dc1}) by
$K_0(x,t|0,\tau)$ and $\psi_0(0,\tau)$, respectively [{\em cf.}
equations (\ref{eq:field_free_propagator}) and
(\ref{eq:wave_function_without_field})], which immediately yields
for $x = 0$
\begin{equation}\label{eq:psi_0_analytic_sol_1}
    {\textstyle \psi_F(0,t)\; \approx\; \phi_F(0,t)\, +\, \sqrt{B}\, e^{-\frac{i}{\hbar}
    E_b t}\; \text{erf}\hspace*{-.0cm}\left(\sqrt{-\frac{i}{\hbar} E_b\,t}\right)\,.}
\end{equation}
Clearly, the second term of the right hand side appeared from the
interaction between the particle and the potential well in the limit
of $F \to 0$. By replacing $\phi_F(0,t)$ here by $\phi_0(0,t)$ as
well and then approximately substituting this $\psi_{F \to
0}(0,\tau)$ into $\psi_F(0,\tau)$ in
(\ref{eq:particular_sol_schroedinger_eq_in_x_dc2_2}), we will be
able to consider a closed expression for $\psi_F(x,t)$ in
(\ref{eq:sol_schroedinger_eq_in_x_dc2}). To this end, we mimic the
procedure for the field-free result in
(\ref{eq:wave_function_without_field}). Then, equation
(\ref{eq:particular_sol_schroedinger_eq_in_x_dc2_2}) is transformed
into \cite{ABR74}
\begin{eqnarray}\label{eq:particular_sol_schroedinger_eq_in_x_dc2_4}
    \hspace*{-1.5cm}&& {\textstyle T_F(x,t,\sigma)}\; =\; {\textstyle \frac{\sqrt{\pi}}{2 \beta_F}
    \left\{e^{-2\,\alpha\,\beta_F}\,
    \int_0^{\sqrt{t}} d s\; \frac{\partial}{\partial s}\,\text{erfc}\hspace*{-.1cm}\left(\frac{\alpha}{s} -
    \beta_F\,s\right)\; -\right.}\n\\
    \hspace*{-1.5cm}&& {\textstyle \left. e^{2\,\alpha\,\beta_F}\,
    \int_0^{\sqrt{t}} d s\; \frac{\partial}{\partial s}\,\text{erfc}\hspace*{-.1cm}\left(\frac{\alpha}{s} +
    \beta_F\,s\right)\right\} \cdot e^{-\frac{i}{\hbar} E_b\,s^2}\; \psi_F(0,t-s^2)\,,}
\end{eqnarray}
where $s = \sqrt{t - \tau}$\,; $\alpha(x) = \frac{|x|}{\sqrt{2 i
\hbar/m}}$\, and\\$\beta_F(x,\sigma) = \sqrt{\frac{i}{\hbar}
\left\{|E_b| - \varepsilon_F(x,\sigma)\right\}}$\,. Now, from
(\ref{eq:airy_representation}) and
(\ref{eq:particular_sol_schroedinger_eq_in_x_dc2_4}) with
$\psi_F(0,t-s^2) \approx \psi_0(0,t-s^2)$, we find that
\begin{equation}\label{eq:sol_schroedinger_eq_in_x_dc3}
    \hspace*{-1.cm}{\textstyle \psi_f(x,t)\; \approx\; \phi_f(x,t)\, +\,
    \frac{\sqrt{B}}{2}\, e^{-\frac{i}{\hbar} E_b t}}\,
    \sum_{k=0}^{\infty} {\textstyle \left.\frac{1}{k!\; 3^k}\;
    T_f^{(3k)}(x,t,\sigma)\right|_{\sigma=0}\;,}
\end{equation}
where the dimensionless quantity $f := \frac{m F}{\hbar^2 B^3}$
represents the relative field strength, and
\begin{eqnarray}
    {\textstyle T_f(x,t,\sigma)} &=& {\textstyle \sqrt{\frac{i |E_b|}{\hbar}}\, \frac{1}{\beta_f(x,\sigma)}}\;
    {\textstyle \left\{e^{-2\,\alpha\,\beta_f}\;\,
    \text{erfc}\hspace*{-.1cm}\left(\frac{\alpha(x)}{\sqrt{t}}\,
    -\, \beta_f(x,\sigma)\,\sqrt{t}\right)\right.}\n\\
    && {\textstyle \left.-\;\, e^{2\,\alpha\,\beta_f}\;\,
    \text{erfc}\hspace*{-.1cm}\left(\frac{\alpha(x)}{\sqrt{t}}\,
    +\, \beta_f(x,\sigma)\,\sqrt{t}\right)\right\}\,.}
\end{eqnarray}
Here, $\beta_f(x,\sigma) = \sqrt{\frac{i |E_b|}{\hbar}} \sqrt{1 - x
B f - \sigma f^{\frac{2}{3}}}$\,. Clearly, equation
(\ref{eq:sol_schroedinger_eq_in_x_dc3}) with $f = 0$ exactly reduces
to its field-free counterpart in
(\ref{eq:wave_function_without_field}).

In the second approximation scheme, we make use of the exponential
decay ansatz for $\psi_F(0,\tau)$ in equation
(\ref{eq:sol_schroedinger_eq_in_x_dc1}),
\begin{equation}\label{eq:psi_0_ionization1}
    {\textstyle \psi_F(0,\tau)\; =\; \sqrt{B}\; e^{-\frac{i}{\hbar} E\,\tau}\,,}
\end{equation}
where the complex-valued energy $E\,=\,E_f - \frac{i}{2}\,\Gamma_f$
with $E_f\,=\,E_b + \Delta_f$; $\Delta_f$ is the level shift.
Equation (\ref{eq:psi_0_ionization1}) with the substitution $\tau =
t - s^2$ easily yields, from equations
(\ref{eq:sol_schroedinger_eq_in_x_dc1}) and (\ref{eq:kernel1_dc})
for $x = 0$,
\begin{equation}\label{eq:psi_0_ionization2}
    {\textstyle \psi_f(0,t)\; =\; \phi_f(0,t)\, +\,
    \sqrt{\frac{2\,i\,\hbar\,B^3\,t}{\pi\,m}}\;
    e^{-\frac{i}{\hbar} E\,t}\; Y_f(t)\,,}
\end{equation}
or alternatively,
\begin{equation}\label{eq:psi_0_ionization2_1}
    {\textstyle \psi_f(0,t)\; =\; \phi_f(0,t)\, \cdot\,
    \left(1\, -\, \sqrt{\frac{2\,i\,\hbar\,B^3\,t}{\pi\,m}}\;
    Y_f(t)\right)^{-1}\,,}
\end{equation}
where $Y_f(t)\,=\,\int_0^1 d z\, e^{-\xi_1\,z^6\, -\, \xi_2\,z^2}$
with $\xi_1\,=\,\frac{f^2\,E_b^3\,t^3}{3\,i\,\hbar^3}$\, and\,
$\xi_2\,=\,\frac{E\,t}{i\,\hbar}$. It has been shown that in the
weak-field limit $|f| \ll 1$ the exponential decay law offers a good
approximation of the exact result for the ionization probability
${\mathcal P}(t)$ \cite{LUD87,KLE94}; further, the semiclassical
value $\Delta_{f,WKB}\,=\,-\frac{5\,\hbar^2 B^2}{8\,m} f^2$ is in
excellent agreement to $\Delta_f$ up to $f \lesssim 0.1$, and
$\Gamma_{f,WKB}\,=\,\frac{\hbar^2 B^2}{m}\,e^{-\frac{2}{3 f}}$ a
good approximation of $\Gamma_f$ for $f \lesssim 1$
\cite{ELK88,KLE94,HAN06}. However, the exponential decay
approximation clearly cannot account for any ripples in the time
evolution of the ionization probability observed from its exact
result. We would like to approximately recover the ripples by using
equations (\ref{eq:psi_0_ionization2}) and
(\ref{eq:psi_0_ionization2_1}) instead of
(\ref{eq:psi_0_ionization1}).

After a lengthy calculation an explicit expression for $Y_f(t)$
reveals itself as
\begin{eqnarray}\label{eq:integral_1}
    Y_f(t) &=& {\textstyle e^{-\xi_1}\,}
    \left\{\,\sum_{k=0}^\infty\, {\textstyle \frac{(-\xi_2)^k}{k!}\,
    \frac{{}_{1}\hspace*{-.05cm}F_1\left(1; \frac{2 k\,+\,7}{6};\,\xi_1\right)}{2
    k\,+\,1}\, +\, 1\, -\, {}_{1}\hspace*{-.05cm}F_1\left(1;
    \frac{7}{6};\,\xi_1\right)}\right.\n\\
    && {\textstyle \left.+\, \frac{6\,\xi_1}{7}\,
    {}_{1}\hspace*{-.05cm}F_1\left(1; \frac{13}{6}; \xi_1\right)\; + \frac{\xi_2^2}{2!\; 5}\,
    \left\{1\, -\, {}_{1}\hspace*{-.05cm}F_1\left(1; \frac{11}{6}; \xi_1\right)\right\}\,
    +\right.}\n\\
    && {\textstyle \left.\frac{3\,\xi_1\,\xi_2^2}{55}\,
    {}_{1}\hspace*{-.05cm}F_1\left(1; \frac{17}{6}; \xi_1\right)\right\}}
\end{eqnarray}
(for the detailed derivation, see Appendix \ref{sec:appendix1}),
which allows us to obtain a closed expression for $\psi_f(0,t)$ in
equation (\ref{eq:psi_0_ionization2}) [{\em or}
(\ref{eq:psi_0_ionization2_1})]. For $\xi_1 = 0$, equation
(\ref{eq:integral_1}) clearly reduces to the field-free result
\begin{equation}\label{eq:error_integral1}
    {\textstyle \int_0^1 d z\; e^{-\xi_2\,z^2}\; =\;} \sum_{k=0}^{\infty}
    {\textstyle \,\frac{(-\xi_2)^k}{k!\, (2k\,+\,1)}\; =\;
    \frac{1}{2}\,{\sqrt\frac{\pi}{\xi_2}}\; \text{erf}(\sqrt{\xi_2})\,.}
\end{equation}
Since there is no guarantee for each expression for $\psi_f(0,t)$ in
(\ref{eq:psi_0_ionization2}) and (\ref{eq:psi_0_ionization2_1}) to
fulfill the normalization condition for the wavefunction, let us
consider their combination for a correct numerical evaluation of
$\psi_f(0,t)$, namely,
\begin{eqnarray}\label{eq:psi_0_ionization2_2}
    {\textstyle \psi_f(0,t)} &=& {\textstyle c\, \left(\phi_f(0,t)\, +\,
    \sqrt{\frac{2\,i\,\hbar\,B^3\,t}{\pi\,m}}\;
    e^{-\frac{i}{\hbar} E\,t}\; Y_f(t)\right)\; +}\\
    && {\textstyle (1\,-\,c)\;\,
    \phi_f(0,t)\,\cdot\,\left(1\, -\, \sqrt{\frac{2\,i\,\hbar\,B^3\,t}{\pi\,m}}\;
    Y_f(t)\right)^{-1}\,,}\n
\end{eqnarray}
where the constant $0\leq c \leq 1$ can be determined by numerical
fitting to the exact result being a numerical solution of the
integral equation (\ref{eq:sol_schroedinger_eq_in_x_dc1}) for $x =
0$. As shown in figures.~\ref{fig:decay} and \ref{fig:shift},
equation~(\ref{eq:psi_0_ionization2_2}) gives the excellent results,
especially in the very short-time region when the field is switched
on so that the decay process is slower than exponential
\cite{KLE94}.

Here, we also find a new and useful integral identity of the Airy
function; with the aid of equations (\ref{eq:airy_function_rel1})
and (\ref{eq:error_integral1}) we easily get
\begin{eqnarray}\label{eq:integral_identity1}
    \hspace*{-.5cm}&&{\textstyle Y_f(t)}\; =\; {\textstyle \int_{-\infty}^{\infty} d\sigma\,
    \text{Ai}(\sigma)\;\int_0^1\, d z\; e^{-\left(\xi_2\, -\,
    \sigma\,(-3\,\xi_1)^{\frac{1}{3}}\right)\,z^2}\; =}\\
    \hspace*{-.5cm}&& {\textstyle \frac{\sqrt{\pi}}{2}\, \int_{-\infty}^{\infty} d\sigma\;
    \text{Ai}(\sigma)\;
    \text{erf}\left(\sqrt{\xi_2\, -\,
    \sigma\,(-3\,\xi_1)^{\frac{1}{3}}}\right)\;
    \frac{1}{\sqrt{\xi_2\, -\,
    \sigma\,(-3\,\xi_1)^{\frac{1}{3}}}}\;.}\n
\end{eqnarray}
At the same time we have \cite{IKI06}
\begin{eqnarray}\label{eq:integral_identity2}
    {\textstyle \int_0^1 d z\, e^{-\xi_1\,z^6}} &=& {\textstyle
    \frac{6}{7}\, M_{\frac{1}{12},\frac{7}{12}}(\xi_1)\;
    e^{-\frac{\xi_1}{2}}\; \xi_1^{-\frac{1}{12}}\; +\;
    e^{-\xi_1}}\n\\
    &=& {\textstyle e^{-\xi_1}\,\left\{\frac{6}{7}\,\xi_1\;
    {}_{1}\hspace*{-.05cm}F_1\left(1; \frac{13}{6}; \xi_1\right)\, +\, 1\right\}\,,}
\end{eqnarray}
where Whittaker's $M$-function\\$M_{\kappa,\mu}(\xi_1)\, =\,
e^{-\frac{\xi_1}{2}}\; \xi_1^{\mu + \frac{1}{2}}\;
{}_{1}\hspace*{-.05cm}F_1\left(\frac{1}{2} + \mu - \kappa ;\, 1 + 2
\mu ;\, \xi_1\right)$, and the confluent hypergeometric
function\\${}_{1}\hspace*{-.05cm}F_1\left(a ;\, b ;\, x\right)\, =\,
\frac{\Gamma(b)}{\Gamma(a)}\, \sum_{k=0}^{\infty}\,
\frac{\Gamma(a\,+\,k)}{\Gamma(b\,+\,k)}\, \frac{x^k}{k!}$\,
\cite{ABS74}. Equation (\ref{eq:integral_identity2}) can easily be
verified by means of the expansion
$e^{-\xi_1\,z^6}\,=\,1\,-\,\xi_1\,z^6\,+\,\cdots$ and then the
comparison of both sides. Now, the comparison between
(\ref{eq:integral_identity1}) with $\xi_2 = 0$ and
(\ref{eq:integral_identity2}) immediately yields the identity
\cite{HNI06}
\begin{eqnarray}\label{eq:airy_identity1}
    && {\textstyle \int_{-\infty}^{\infty} \frac{d\sigma}{\sqrt{\sigma}}\;
    \text{Ai}(\sigma)\;
    \text{erf}(\chi\,\sqrt{\sigma})}\n\\
    &=& {\textstyle \frac{2\,\chi}{\sqrt{\pi}}\; e^{-\xi_1}\, \left\{\frac{6}{7}\,\xi_1\;
    {}_{1}\hspace*{-.05cm}F_1\left(1; \frac{13}{6}; \xi_1\right)\, +\, 1\right\}\,,}
\end{eqnarray}
where $\xi_1 = \chi^6/3$.

\section{Time-Evolution of the probability density at the potential center}
\label{sec:ionization_probability}
It is still not easy to deal with a closed expression in
(\ref{eq:ionization_prob_pre1}) for the numerical evaluation of the
ionization probability ${\mathcal{P}}(t) = 1 -
|\<\psi_b|\psi_F(t)\>|^2$. Considering the shape of $\psi_b(x) =
\sqrt{B}\,e^{-B\,|x|}$ for $B$ large enough ({\em or} $|f| \ll 1$
for a given field strength $F$), it would be no harm to adopt for
the qualitative study of ${\mathcal{P}}(t)$ the approximation
\begin{equation}\label{eq:approximation_delta1}
    \lim_{B \to \infty} \sqrt{B}\, e^{-B\,|x|}\;\; \longrightarrow\;\;
    \delta(x)\,=\,\lim_{n \to \infty}
    \sqrt{\frac{n}{\pi}}\; e^{-n\,x^2}\,,
\end{equation}
which allows us to use $1 - |\psi_f(0,t)|^2$ as a simple alternative
to ${\mathcal{P}}(t)$. See figure~\ref{fig:abssqr} for the numerical
evaluation of $|\psi_f(0,t)|^2$ for various values of $f$; the
closed expression in (\ref{eq:psi_0_ionization2_2}) with
(\ref{eq:integral_1}) gives results in good agreement to the exact
ones \cite{KIM05}. Here we observe the slack ripples in the time
evolution obtained from equation (\ref{eq:psi_0_ionization2_2}). The
origin of the ripples has been discussed in \cite{ELK88} and
\cite{REI85}; the initial bound state $\varphi_b(p)$ in equation
(\ref{eq:bound_state_momentum_rep}) is symmetric around $p = 0$. By
applying the field, this symmetry breaks down in such a way that the
motion of the particles in one direction is just accelerated so that
they will easily leave the potential well. On the other hand, the
other direction gets slowed down until the particles stop, and then
they reverse their direction of motion so that the particles again
approach the potential well at $x = 0$. Then they are partially
reflected from the potential well. This process is repeated until
all particles will completely leave the potential well. From the
results shown in figures~\ref{fig:decay}-\ref{fig:abssqr} we may say
that $\psi_f(0,\tau)$ in (\ref{eq:psi_0_ionization2_2}) leads to a
good approximation of the exact result for $\psi_f(x,t)$ in the
integral equation (\ref{eq:sol_schroedinger_eq_in_x_dc1}).

\section{Conclusion}\label{sec:conclusion}
In summary, we studied the time evolution of a particle bound by an
attractive one-dimensional delta-function potential (at $x = 0$)
({\em or} an ultrathin quantum well) when a uniform electrostatic
field is applied. Thus far, no analytically solvable model of field
emission has been known. We obtained explicit expressions for the
time-dependent wavefunction $\psi_f(x,t)$ and the ionization ({\em
or} bound state) probability ${\mathcal{P}}(t)$, respectively, in
the weak-field limit, especially that for $\psi_f(0,t)$ [see
equations (\ref{eq:psi_0_ionization2}),
(\ref{eq:psi_0_ionization2_1}) and their combination
(\ref{eq:psi_0_ionization2_2}) with (\ref{eq:integral_1})] which is
a key element to the evaluation of $\psi_f(x,t)$ and
${\mathcal{P}}(t)$. This explicit expression for $\psi_f(0,t)$ was
shown to be a much better approximation of the exact result than its
counterpart obtained from the exponential decay approximation [see
equation (\ref{eq:psi_0_ionization1})] in that, e.g., the resulting
probability density $|\psi_f(0,t)|^2$ as a simple alternative to
${\mathcal{P}}(t)$ can easily be numerically evaluated and is in
excellent agreement to the exact result with the ripples, whereas no
ripples can be observed from the exponential decay approximation. It
is further suggested that even for the strong-field limit our result
for $\psi_f(0,t)$ would be a better approximation {\em on the
average} to the (highly oscillatory) exact one than the result from
the exponential decay law which has been shown to be a good
approximation on the average in the strong-field limit \cite{ELK88}.
In studying this subject, we also found an interesting integral
identity of the Airy function. Next, we will explore the analytical
expressions for various (time-dependent) quantities in a
delta-function potential system
$V(x)\,=\,\sum_{\lambda}\,V_{\lambda}\; \delta(x - x_{\lambda})$,
where $\lambda = 0, 1, 2,\,\cdots\,N$.\vspace*{.5cm}

\section*{Acknowledgments}
The author would like to thank G.J. Iafrate for the stimulating
discussion.

\appendix*\section{: A mathematical supplement - derivation of
equation (\ref{eq:integral_1})}\label{sec:appendix1}
We would like to evaluate the integral $\int_0^1 d z\;
e^{-\xi_1\,z^6\, -\, \xi_2\,z^2}\, =\, A_0(\xi_1, \xi_2)\, +\,
A_1(\xi_1, \xi_2)\, +\, A_2(\xi_1, \xi_2)$, where
\begin{equation}\label{eq:apendix_eq1}
    {\textstyle A_k(\xi_1, \xi_2)\; =\;} \sum_{n = 0}^{\infty}\,
    {\textstyle \frac{(-\xi_2)^{3 n\,+\,k}}{(3 n\,+\,k)!}\,\int_0^1\, d z\;
    e^{-\xi_1\,z^6}\; z^{6 n\,+\,2 k}\;.}
\end{equation}
Along the same line with eq. (\ref{eq:integral_identity2}), we can
obtain
\begin{equation}\label{eq:appendix_integral0}
    {\textstyle \int_0^1\, d z\;
    e^{-\xi_1\,z^6}\; z^{6 m}\; =\; e^{-\xi_1}\,
    \frac{{}_{1}\hspace*{-.05cm}F_1\left(1 ;\, m\,+\,\frac{7}{6} ;\, \xi_1\right)}{6 m\,+\,1}}
\end{equation}
with $m = 1, 2, \cdots$, which immediately leads to
\begin{eqnarray}\label{eq:integral_sum1}
    {\textstyle A_0(\xi_1, \xi_2)} &=& {\textstyle e^{-\xi_1}\,}
    \left\{{\textstyle 1\,+\,\frac{6\,\xi_1}{7}\,
    {}_{1}\hspace*{-.05cm}F_1\left(1 ;\, \frac{13}{6} ;\, \xi_1\right)\,
    +\,}\right.\n\\
    && \left.\sum_{m = 1}^{\infty}\, {\textstyle \frac{(-\xi_2)^{3 m}}{(3 m)!}\;
    \frac{{}_{1}\hspace*{-.05cm}F_1\left(1 ;\, m\,+\,\frac{7}{6} ;\, \xi_1\right)}{6 m\,+\,1}}\right\}\,.
\end{eqnarray}
Similarly, we get
\begin{equation}\label{eq:appendix_integral1}
    {\textstyle \int_0^1\, d z\;
    e^{-\xi_1\,z^6}\; z^{6 n\,+\,2}\; =\; e^{-\xi_1}\,
    \frac{{}_{1}\hspace*{-.05cm}F_1\left(1 ;\, n\,+\,\frac{3}{2} ;\, \xi_1\right)}{6 n\,+\,3}\,,}
\end{equation}
and then
\begin{equation}\label{eq:integral_sum2}
    {\textstyle A_1(\xi_1, \xi_2)\; =\; e^{-\xi_1}\,}\, \sum_{n = 0}^{\infty}\,
    {\textstyle \frac{(-\xi_2)^{3 n + 1}}{(3 n\,+\,1)!}\;
    \frac{{}_{1}\hspace*{-.05cm}F_1\left(1 ;\,n\,+\,\frac{3}{2} ;\, \xi_1\right)}{6 n\,+\,3}\,.}
\end{equation}
Also, it follows that
\begin{eqnarray}\label{eq:integral_sum3}
    && {\textstyle \int_0^1\, d z\;
    e^{-\xi_1\,z^6}\; z^{4}}\; =\; {\textstyle 2\, e^{-\xi_1}\,
    \left\{\frac{1}{10}\, +\, \frac{3\,\xi_1}{55}\,
    {}_{1}\hspace*{-.05cm}F_1\left(1 ;\, \frac{17}{6} ;\,
    \xi_1\right)\right\}\;,}\n\\
    && {\textstyle \int_0^1\, d z\;
    e^{-\xi_1\,z^6}\; z^{6 m\,+\,4}}\; =\; {\textstyle e^{-\xi_1}\,
    \frac{{}_{1}\hspace*{-.05cm}F_1\left(1 ;\, m\,+\,\frac{11}{6} ;\, \xi_1\right)}{6 m\,+\,5}}\;,\n\\
    && {\textstyle A_2(\xi_1, \xi_2)}\; =\; {\textstyle e^{-\xi_1}\;
    \xi_2^2}\, \left\{{\textstyle \frac{1}{10}\,+\,\frac{3\,\xi_1}{55}\,
    {}_{1}\hspace*{-.05cm}F_1\left(1 ;\, \frac{17}{6} ;\, \xi_1\right)\; +}\right.\n\\
    && \hspace*{2cm}\left.\sum_{m = 1}^{\infty}\, {\textstyle \frac{(-\xi_2)^{3 m}}{(3 m\,+\,2)!}\;
    \frac{{}_{1}\hspace*{-.05cm}F_1\left(1 ;\, m\,+\,\frac{11}{6} ;\, \xi_1\right)}{6 m\,+\,5}}\right\}\,.
\end{eqnarray}
From equations (\ref{eq:integral_sum1}), (\ref{eq:integral_sum2}),
and (\ref{eq:integral_sum3}) we easily arrive at the expression in
(\ref{eq:integral_1}).

\section*{Figure captions}

\begin{figure}
    \hspace*{-3cm}
    \resizebox{0.75\textwidth}{!}{%
    \includegraphics{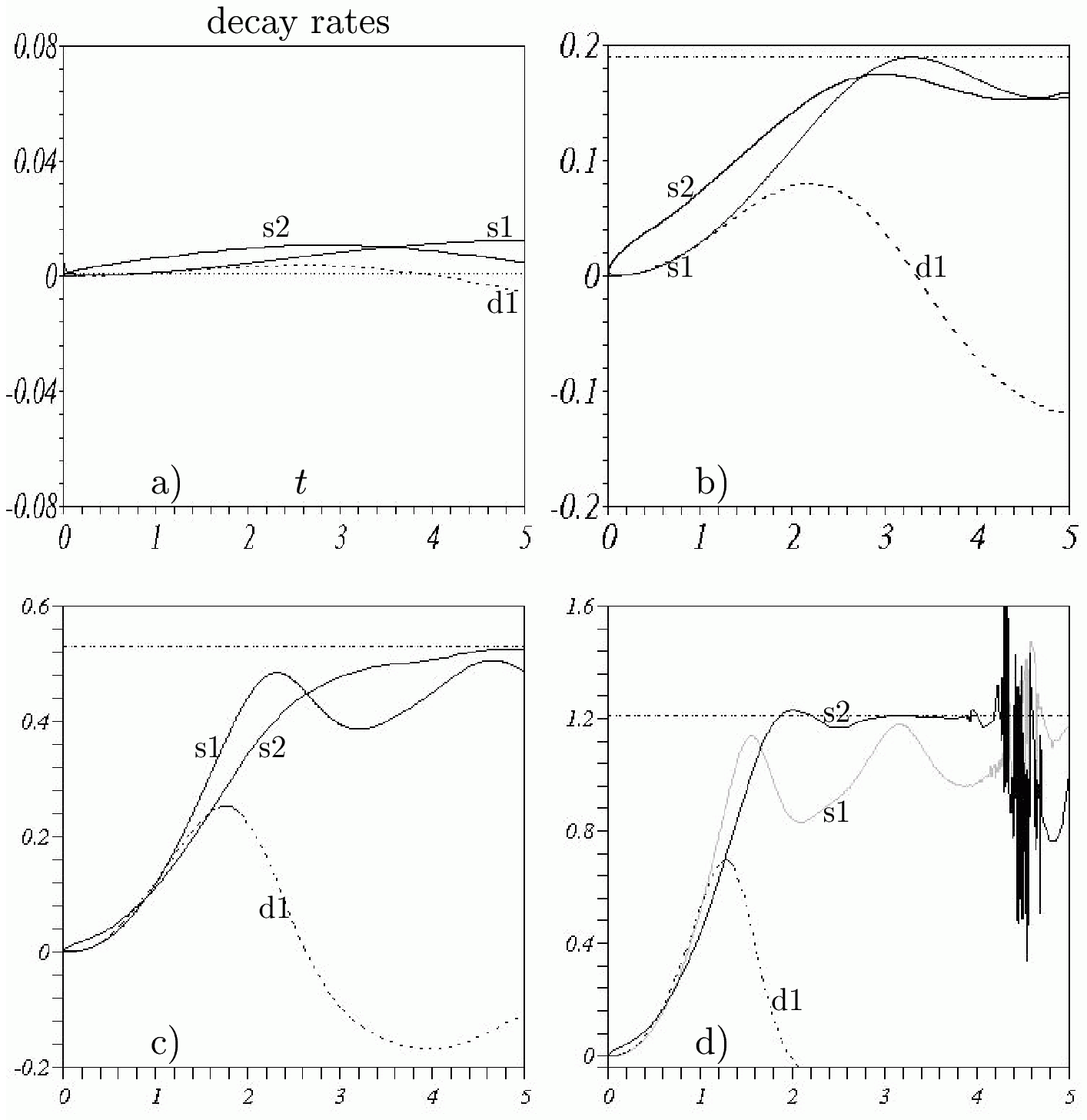}}
    \caption{\label{fig:decay}}
\end{figure}

Figure~\ref{fig:decay}: decay rates $\Gamma_f(t)$ versus time $t$,
where $\Gamma_f(t)$ is obtained from the wavefunction $\psi_f(0,t) =
e^{-\frac{i}{\hbar} {\mathcal{A}}(t)}$ with ${\mathcal{A}}(t) =
\left(E_b + \Delta_f(t) - \frac{i}{2} \Gamma_f(t)\right) t$. Here
the bound state energy $E_b = -\frac{\hbar^2 B^2}{2 m}$, and we use
$\hbar = m = B = 1$ for plots. Two solid lines are {\bf{s1}} from
the exact result and {\bf s2} from equation
(\ref{eq:psi_0_ionization2_2}) while two dot lines are {\bf d1} from
equation (\ref{eq:psi_0_analytic_sol_1}) and {\bf d2} (straight
line) from the exponential decay law. a) for $f = 0.1$, $c = 1$ for
{\bf s2}, and $\Gamma_f = 0.0010$ for {\bf d2} ; b) for $f = 0.5$,
$c = 0.65$ for {\bf s2}, and $\Gamma_f = 0.1896$ for {\bf d2} ; c)
for $f = 1$, and $c = 0.45$ for {\bf s2}, and $\Gamma_f = 0.52916$
for {\bf d2} ; d) for $f = 2$, $c = 0.45$ for {\bf s2}, and
$\Gamma_f = 1.2115$ for {\bf d2}.\vspace*{.3cm}

\begin{figure}
    \hspace*{-4cm}
    \resizebox{0.75\textwidth}{!}{%
    \includegraphics{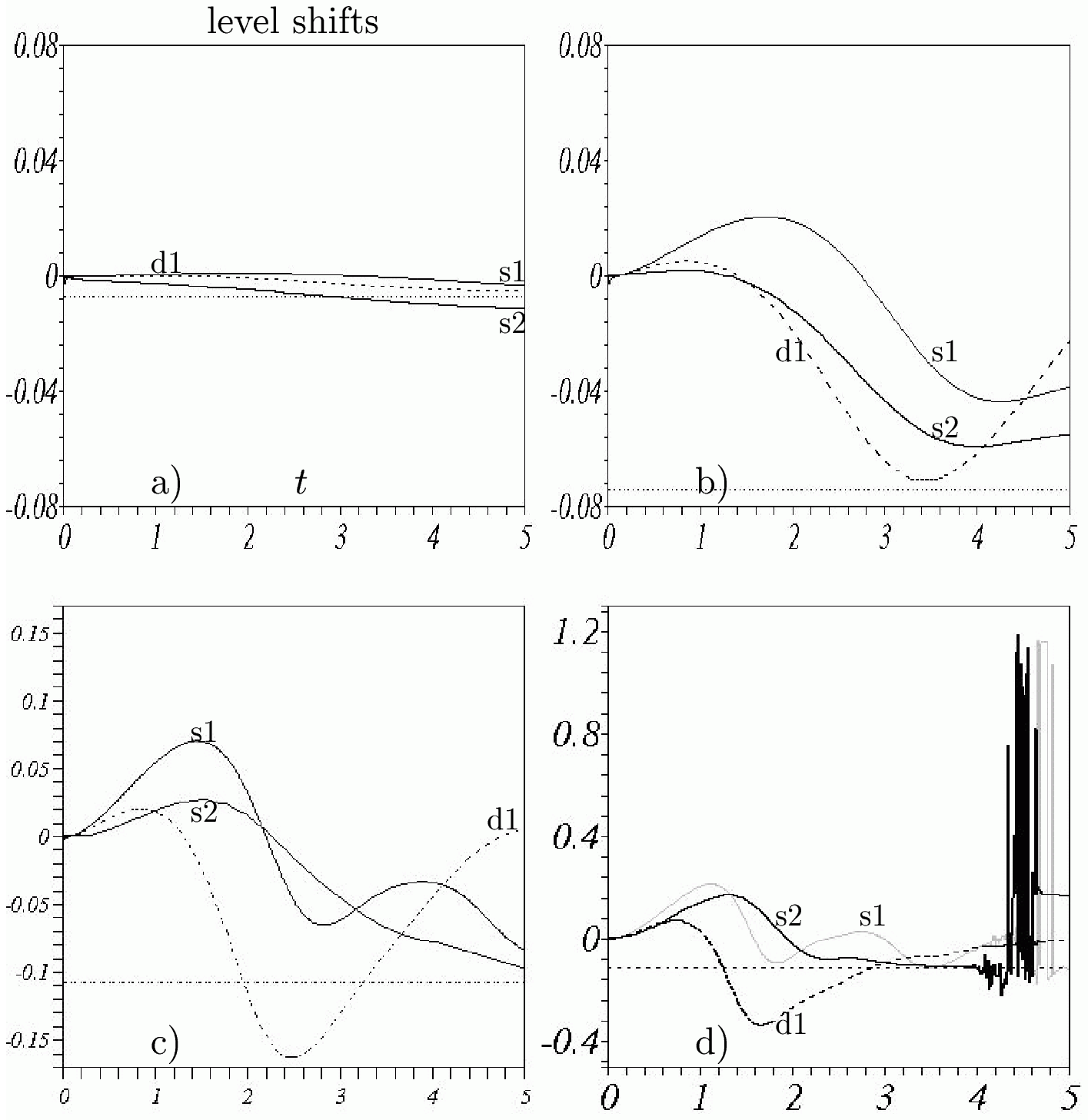}}
    \caption{\label{fig:shift}}
\end{figure}

\noindent{Figure~\ref{fig:shift}:} level shifts $\Delta_f(t)$ versus
time $t$, where $\Delta_f(t)$ is obtained from $\psi_f(0,t)$ as
$\Gamma_f(t)$ is. a) for $f = 0.1$ and $\Delta_f = -0.0072$ for {\bf
d2} ; b) for $f = 0.5$ and $\Delta_f = -0.0738$ for {\bf d2} ; c)
for $f = 1$ and $\Delta_f = -0.10722$ for {\bf d2} ; d) for $f = 2$
and $\Delta_f = -0.11235$ for {\bf d2}. Other parameters are the
same as for figure~\ref{fig:decay}.\vspace*{.3cm}

\begin{figure}
    \hspace*{-3cm}
    \resizebox{0.75\textwidth}{!}{%
    \includegraphics{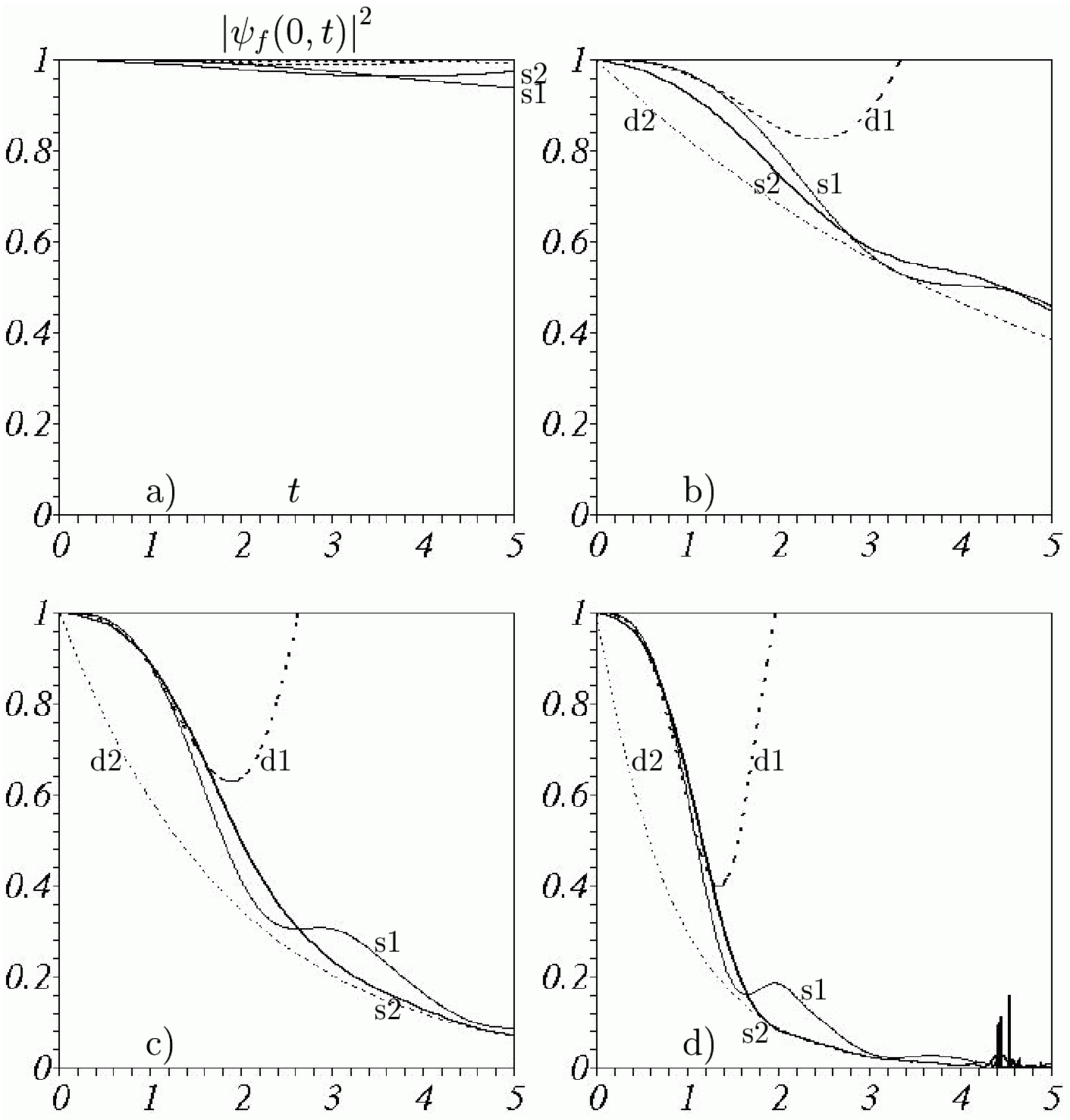}}
    \caption{\label{fig:abssqr}}
\end{figure}

\noindent{Figure~\ref{fig:abssqr}:} $\left|\psi_f(0,t)\right|^2$
versus time $t$. All parameters are the same as for figures
\ref{fig:decay} and \ref{fig:shift}. Note that {\bf d1} breaks down
for $f = 0.5,\,1,\,2$, i.e., $\left|\psi_f(0,t)\right|^2$ becomes
greater than $1$.

\begin{thebibliography}{1}
\bibitem{ABR74} Here, we used the identity $\int dy\,e^{-a^2 y^2 - b^2/y^2}\,
=\, \frac{\sqrt{\pi}}{4\,a}\, \{e^{2 a b}\, \text{erf}(a y + b/y)\,
+\, e^{-2 a b}\, \text{erf}(a y - b/y)\}\, +\, \text{const}\, =\,
\frac{\sqrt{\pi}}{4\,a}\, \{-e^{2 a b}\, \text{erfc}(a y + b/y)\,
+\, e^{-2 a b}\, \text{erfc}(-a y + b/y)\}\, +\, \text{const}$,
which can be found, for instance, in Ref. \cite{ABS74}.
\bibitem{KIM06} From the Airy-function identity ${\textstyle \int_{-\infty}^{\infty}\,
d\sigma\, \text{Ai}(\sigma)\; e^{\sigma u}\, =}$ ${\textstyle
e^{\frac{1}{3} u^3}}$, we easily see that three different values for
$u$, namely, $u_1 = i \eta$, $u_2 = i\,e^{-i\frac{2\pi}{3}} \eta$,
and $u_3 = i\,e^{-i\frac{4\pi}{3}} \eta$ actually satisfy equation
(\ref{eq:airy_function_rel1}). Here, we pick up $u_1$.
\bibitem{HAN06} The exact values of $\Gamma_f$ and $\Delta_f$ for any $f$ can be
obtained in the scheme of exponential decay law; refer to, e.g.,
equation (4.32) with (4.10) and (4.33) of Ref. \cite{KLE94}.
\bibitem{IKI06} Equation (\ref{eq:integral_identity2}) can be found, e.g., by using the Maple.
\bibitem{HNI06} The identity in (\ref{eq:airy_identity1}) was not
given in Ref. \cite{VAL04} which contains many accumulated results
of analytical properties of the Airy function.
\end{thebibliography}
\end{document}